\def\prg#1{\medskip{\bf #1}}
        
               \def\pd{\partial}
\def\dis{\displaystyle}          
\def\grl{{GR$_\Lambda$}}         \def\vsm{\vspace{-10pt}}
  
\def\ads3{AdS$_3$}               \def\cs{{\sc cs}}
\def\Leff{\hbox{$\mit\L_{\hspace{.6pt}\rm eff}\,$}}
\def\bull{\raise.25ex\hbox{\vrule height.8ex width.8ex}}

\def\cL{{\cal L}}     \def\cM{{\cal M }}    \def\cK{{\cal K}}
\def\cO{{\cal O}}     \def\cE{{\cal E}}     \def\cH{{\cal H}}
\def\tR{{\tilde R}}   \def\tG{{\tilde G}}   \def\hcO{\hat{\cal O}}
\def\hcH{\hat{\cH}}   \def\tI{{\tilde I}}   
\def\bcH{{\bar\cH}}   \def\bcK{{\bar\cK}}   \def\bD{{\bar D}}
           \def\cC{{\cal C}}
\def\Ic{{I_{\rm c}}}  \def\tIc{{\tilde\Ic}} \def\bI{{\bar I}}

\def\G{\Gamma}                \def\L{{\mit\Lambda}}
\def\a{\alpha}        \def\b{\beta}         \def\g{\gamma}
\def\d{\delta}        \def\m{\mu}           \def\n{\nu}
\def\th{\theta}               \def\l{\lambda}
\def\vphi{\varphi}    \def\ve{\varepsilon}  
\def\r{\rho}          \def\Om{\Omega}       \def\om{\omega}
                  \def\tom{{\tilde\omega}}
\def\nn{\nonumber}
\def\be{\begin{equation}}             \def\ee{\end{equation}}
\def\ba#1{\begin{array}{#1}}          \def\ea{\end{array}}
\def\bea{\begin{eqnarray} }           \def\eea{\end{eqnarray} }
\def\beann{\begin{eqnarray*} }        \def\eeann{\end{eqnarray*} }
\def\beal{\begin{eqalign}}            \def\eeal{\end{eqalign}}
\def\lab#1{\label{eq:#1}}             \def\eq#1{(\ref{eq:#1})}
\def\bsubeq{\begin{mathletters}}      \def\esubeq{\end{mathletters}}
\def\bitem{\begin{itemize}}           \def\eitem{\end{itemize}}
\documentstyle[aps,preprint,eqsecnum]{revtex}

\begin{document}
\tighten


\title{Black hole entropy in 3D gravity with torsion}

\author{M.\ Blagojevi\'c and B. Cvetkovi\'c\thanks{Email addresses:
                mb@phy.bg.ac.yu, cbranislav@phy.bg.ac.yu}\\
Institute of Physics, P.O.Box 57, 11001 Belgrade, Serbia}
\maketitle
\begin{abstract}
The role of torsion in three-dimensional quantum gravity is
investigated by studying the partition function of the Euclidean
theory in Riemann-Cartan spacetime. The entropy of the black hole
with torsion is found to differ from the standard Bekenstein-Hawking
result, but its form is in complete agreement with the first law of
black hole thermodynamics. 
\end{abstract}

\section{Introduction}

In our attempts to properly understand basic features of the
gravitational dynamics at both classical and quantum level, black holes
are often used as an arena for testing new ideas. In the early 1970s,
Bekenstein \cite{1} and Hawking \cite{2} discovered that black holes
are thermodynamic objects, with characteristic temperatures and
entropies. An intensive study of these concepts has led us to conclude
that they are closely related to the quantum nature of gravity. In this
regard, the discovery of the BTZ black hole in three-dimensional (3D)
gravity was of particular importance, as it allowed us to investigate
these issues in a substantially simpler context \cite{3}.

Following the traditional approach based on general relativity (GR),
3D gravity has been studied mainly in the realm of Riemannian
geometry, leading to a number of outstanding results
\cite{4,5,6,7,8,9,10}. However, it has, already from the 1960s, been
well-known that there is a more natural, gauge-theoretic conception
of gravity based on {\it Riemann-Cartan geometry\/}, which contains
both the {\it curvature\/} and the {\it torsion\/} of spacetime as
basic elements of the gravitational dynamics (see, e.g.
\cite{11,12}). The application of these ideas to 3D gravity started
in the 1990s by Mielke, Baekler and Hehl \cite{13}, see also
\cite{14}. Recent developments in this direction led to several
interesting conclusions: (a) the Mielke-Baekler model of 3D gravity
with torsion possesses the black hole solution, (b) it can be
formulated as a Chern-Simons gauge theory, and (c) suitable
asymptotic conditions generate the asymptotic conformal symmetry,
described by two independent Virasoro algebras with different central
charges \cite{15,16,17,18}. In the present paper, we continue our
study of 3D gravity with torsion by investigating the important
concept of black hole entropy. Using the Euclidean formulation of the
Mielke-Baekler model, we found a new expression for the black hole
entropy, and examined its consistency with the first law of black
hole thermodynamics.

The paper is organized as follows. In Sect. II, we discuss basic
aspects of the Euclidean 3D gravity with torsion, defined by the
Mielke-Baekler action \cite{13}. In Sect. III, we recover the related
black hole solution, which is of particular importance for
thermodynamic considerations. In Sect. IV, we derive the canonical
expressions for energy and angular momentum of the Euclidean black
hole with torsion. Sect. V contains basic results of the paper.
First, assuming that the black hole manifold contains an inner
boundary at the horizon (as explained in subsection V.A), we derive a
new expression for the entropy of the black hole with torsion. Beside
the Bekenstein-Hawking term, it contains an additional contribution,
which depends on both the strength of torsion and the position of
``inner horizon". For vanishing torsion, the entropy reduces to the
form found earlier by Solodukhin in the context of Riemannian
geometry, but with a Chern-Simons term in the action \cite{19}. Then,
using our results for the black hole energy, angular momentum and
entropy, we prove the validity of the first law of black hole
thermodynamics. Thus, torsion is shown to be in complete agreement
with the first law. Finally, Sect. VI is devoted to concluding
remarks, while appendices contain some technical details on Euclidean
continuation and the hyperbolic geometry in 3D.

In our conventions, Minkowskian 3D gravity with torsion is defined by
an action $I_M$ in Riemann-Cartan spacetime with signature
$\eta^M=(+,-,-)$ \cite{18}. The analytic continuation of the theory
is determined by another action $\bI$, such that $iI_M\mapsto\bI$, in
spacetime with $\bar\eta=(-,-,-)$. Although $\bI$ is the object of
our prime interest, technically, we base our exposition on the
standard Euclidean formalism defined by $iI_M\mapsto-I_E$ and
$\eta^E=(+,+,+)$. Our conventions are given by the following rules:
index {\small $E$} is omitted for simplicity, the Greek indices refer
to the coordinate frame, the Latin indices refer to the tangent
frame; the middle alphabet letters $(i,j,k,\dots;\m,\n,\l,\dots)$ run
over $0,1,2$, while the first letters of the Greek alphabet
$(\a,\b,\g,\dots)$ run over $1,2$; $\eta_{ij}=$ diag$\,(1,1,1)$ are
the tangent frame components of the metric; totally antisymmetric
tensor $\ve^{ijk}$ and the related tensor density $\ve^{\m\n\r}$ are
both normalized by $\ve^{012}=+1$.

\section{Euclidean gravity with torsion}

Following the analogy with Poincar\'e gauge theory \cite{11,12},
Euclidean gravity with torsion in 3D can be formulated as a gauge
theory of the Euclidean group $E(3)=ISO(3)$ (EGT for short), the
analytic continuation of the Poincar\'e group $P(3)=ISO(1,2)$. The
underlying geometric structure is described by Riemann-Cartan space.

\prg{EGT in brief.} Basic gravitational variables in EGT are the
triad field $b^i$ and the spin connection $A^{ij}=-A^{ji}$ (1-forms).
The corresponding field strengths are the torsion and the curvature:
$T^i=db ^i+A^i{_m}\wedge b^m$, $R^{ij}=dA^{ij}+A^i{_m}\wedge A^{mj}$
(2-forms). Gauge symmetries of the theory are local translations and
local rotations, parametrized by $\xi^\m$ and $\ve^{ij}$.

In 3D, we can simplify the notation by introducing
$$
A^{ij}=-\ve^{ijk}\om_k\, ,\quad R^{ij}=-\ve^{ijk}R_k\, ,
\quad \ve^{ij}=-\ve^{ijk}\th_k\, .
$$
In local coordinates $x^\m$, we can write $b^i=b^i{_\m}dx^\m$,
$\om^i=\om^i{_\m}dx^\m$, the field strengths are
\bea
&&T^i=db^i+\ve^i{}_{jk}\om^j\wedge b^k = D b^i\, ,         \nn\\
&&R^i=d\om^i+\frac{1}{2}\,\ve^i{}_{jk}\om^j\wedge\om^k\, , \lab{2.1}
\eea
and gauge transformations take the form
\bea
\d_0 b^i{_\m}&=&
   -\ve^i{}_{jk}b^j{}_{\m}\th^k-(\pd_\m\xi^\r)b^i{_\r}
   -\xi^\r\pd_\r b^i{}_\m \, ,                             \nn\\
\d_0\om^i{_\m}&=&-\nabla_\m\th^i-(\pd_\m\xi^\r)\om^i{_\r}
   -\xi^\r\pd_\r\om^i{}_\m \, ,                            \lab{2.2}
\eea
where $\nabla_\m\th^i=\pd_\m\th^i+\ve^i{}_{jk}\om^j{_\m}\th^k$. The
covariant derivative $\nabla\equiv dx^\m\nabla_\m$ acts on a general
tangent-frame spinor in accordance with its spinorial structure,
while $D X=\nabla\wedge X$ is the covariant exterior derivative of a
form.

The metric structure of EGT is defined by
$$
g=\eta_{ij}b^i\otimes b^j\equiv g_{\m\n}dx^\m\otimes dx^\n,
\quad \eta_{ij}=\mbox{diag}\,(1,1,1)\, .
$$
Although metric and connection on an arbitrary manifold can be
specified as independent fields, in EGT they are related to each
other by the {\it metricity condition\/}: $\nabla g=0$. Consequently,
the geometric structure of EGT corresponds to {\it Riemann-Cartan
geometry\/}.

We display here a useful EGT identity:
\be
\om^i\equiv\tom^i+K^i\, ,                                  \lab{2.3}
\ee
where $\tom^i$ is the Levi-Civita (Riemannian) connection, and $K^i$
is the contortion 1-form, defined implicitly by
$T^i=\ve^i{}_{mn}K^m\wedge b^n$.

\prg{Topological action.} General gravitational dynamics is
determined by Lagrangians which are at most quadratic in field
strengths. Omitting the quadratic terms, we get the {\it topological\/}
model for 3D gravity, proposed by Mielke and Baekler \cite{13}:
\bsubeq\lab{2.4}
\be
I=aI_1+\L I_2+\a_3I_3+\a_4I_4+I_M\, ,                      \lab{2.4a}
\ee
where $I_M$ is a matter contribution, and
\bea
&&I_1= 2\int b^i\wedge R_i\, ,                             \nn\\
&&I_2=-\frac{1}{3}\,\int\ve_{ijk}b^i\wedge b^j\wedge b^k\,,\nn\\
&&I_3=\int\left(\om^i\wedge d\om_i
  +\frac{1}{3}\ve_{ijk}\om^i\wedge\om^j\wedge\om^k\right)\,,\nn\\
&&I_4=\int b^i\wedge T_i\, .                               \lab{2.4b}
\eea
\esubeq
The first term, with $a=1/16\pi G$, is the usual Einstein-Cartan
action, the second term is a cosmological term, $I_3$ is the
Chern-Simons action for the spin connection, and $I_4$ is a torsion
counterpart of $I_1$. The Mielke-Baekler model is a natural
generalization of GR with a cosmological constant (\grl).

\prg{The vacuum field equations.} Variation of the action with
respect to $b^i$ and $\om^i$ yields the gravitational field equations.
Dynamical properties in the region {\it outside\/} the gravitational
sources are determined by the field equations {\it in vacuum\/}:
\bea
&&2aR_i+2\a_4 T_i-\L\ve_{ijk}b^j\wedge b^k=0\, ,           \nn\\
&&2aT_i+2\a_3 R_i+\a_4\ve_{ijk}b^j\wedge b^k=0\, .         \nn
\eea
In the sector $\a_3\a_4-a^2\ne 0$, these equations take the simple
form \cite{*}
\bea
2T^i=p\ve^i{}_{jk}\,b^j\wedge b^k\, ,\qquad
2R^i=q\ve^i{}_{jk}\,b^j\wedge b^k\, ,                      \lab{2.5}
\eea
with
$$
p=\frac{\a_3\L+\a_4 a}{\a_3\a_4-a^2}\, ,\qquad
q=-\frac{(\a_4)^2+a\L}{\a_3\a_4-a^2}\, .
$$
Thus, vacuum solutions are characterized by constant torsion and
constant curvature. For $p=0$, the vacuum geometry is Riemannian,
while for $q=0$, it becomes teleparallel. Note that $p$ and $q$
satisfy the following identities:
$$
aq+\a_4 p-\L=0\, ,\qquad ap+\a_3q+\a_4=0\, .
$$

In Riemann-Cartan spacetime, one can use the identity \eq{2.3} to
express the curvature $R^i=R^i(\om)$ in terms of its {\it
Riemannian\/} piece $\tR^i=R^i(\tom)$ and the contortion:
$$
R^i=\tR^i+DK^i-\frac{1}{2}\ve^{imn}K_m\wedge K_n\, .
$$
This identity, combined with the relation $K^i=p\,b^i/2$, which
follows from the field equations \eq{2.5}, leads to
\bsubeq\lab{2.6}
\be
2\tR^i=\Leff\ve^i{}_{jk}\,b^j\wedge b^k\, ,\qquad
\Leff\equiv q-\frac{1}{4}p^2\, ,                           \lab{2.6a}
\ee
or equivalently:
\be
\tR^{ij}=-\Leff b^i\wedge b^j\, ,                          \lab{2.6b}
\ee
\esubeq
where $\Leff$ is the effective cosmological constant. Using the
Riemannian terminology, we can say that our spacetime is maximally
symmetric, in the sense that its metric has maximal number of Killing
vectors. In what follows, our attention will be focused on the model
\eq{2.4} with $\a_3\a_4-a^2\ne 0$, and with {\it positive\/} $\Leff$
(Euclidean anti-de Sitter sector):
\be
\Leff\equiv\frac{1}{\ell^2}>0 \, .                         \lab{2.7}
\ee
The corresponding Riemannian scalar curvature is negative:
$\tR=-6\Leff$.

\section{The black hole solution}

For positive $\Leff$, the field equation \eq{2.6} has a well-known
solution for the metric---the Euclidean BTZ black hole. In spite of
its dynamical complexity, this solution enables a simple approach to
the gravitational thermodynamics, based on the observation that the
Euclidean action at the black hole contains non-trivial thermodynamic
information \cite{8,9,20,22}. The Euclidean black hole metric in
Schwarzschild-like coordinates reads
\bea
&&ds^2=N^2dt^2+N^{-2}dr^2+r^2(d\vphi+N_\vphi dt)^2\, ,     \lab{3.1}\\
&&N^2=\left(-8Gm+\frac{r^2}{\ell^2}-\frac{16G^2J^2}{r^2}\right),
  \quad N_\vphi=-\frac{4GJ}{r^2}\, .                       \nn
\eea
Since the Riemannian curvature of the solution is negative,
$\tR=-6/\ell^2$, $\Leff$ is positive. The zeros of $N^2$, $r_+$ and
$r_-\equiv-i\r_-$, are related to the black hole parameters by
relations
$$
8G\ell^2 m=r^2_+-\r_-^2\, ,\qquad 4G\ell J=r_+\r_-\, .
$$
The Euclidean metric \eq{3.1} is obtained from the corresponding
Minkowskian expression \cite{18} by the process of analytic
continuation, described in Appendix A. In the Euclidean sector, both
$\vphi$ {\it and\/} $t$ are taken to be periodic (Appendix B),
$$
0\le\vphi<2\pi\, ,\qquad 0\le t<\b\, ,\qquad
\b\equiv\frac{2\pi\ell^2 r_+}{r_+^2+\r_-^2}\, ,
$$
while the radial coordinate $r$ is in the range $ r_+\le r<\infty$.
Topologically, any surface with constant $r$ is an ordinary 2-torus,
parametrized by $\vphi$ and $t$, and the whole black hole manifold
$\cM$ is a {\it solid torus\/}. The black hole horizon $r=r_+$ is a
circle at the core of the solid torus, so that the manifold does not
contain the region $r<r_+$, corresponding to the inner part of the
Minkowskian black hole. It is clear that $\cM$ has the (asymptotic)
boundary located at spatial infinity $r\to\infty$. Later, in subsection
V.A, we shall reconsider arguments in favor of the assumption that the
horizon should be regarded as an additional, inner boundary of $\cM$
\cite{8,21,22}.

The horizon is a one-dimensional subspace $r=r_+$ with the metric
$ds^2=\ell^2d\psi^2$, where $\psi=r_+\vphi/\ell-\r_- t/\ell^2$. The
``area" of the horizon is $2\pi r_+$. For later convenience, we
introduce the quantity
$$
\Om=N_\vphi(r_+)=-\frac{\r_-}{\ell r_+}\, ,
$$
which defines the black hole angular velocity.

Starting with the BTZ metric \eq{3.1}, we construct the black hole
with torsion in the following two steps. First,  we choose $b^i$ to
have the simple, ``diagonal" form:
\bsubeq\lab{3.2}
\bea
&&b^0=Ndt\, ,\qquad b^1=N^{-1}dr\, ,                       \nn\\
&&b^2=r\left(d\vphi+N_\vphi dt\right)\, .                  \lab{3.2a}
\eea
Then, we combine the field equation $K^i=p\,b^i/2$ with the identity
\eq{2.3}, and obtain the connection:
\be
\om^i=\tom^i+\frac{p}{2}\,b^i\, ,                          \lab{3.2b}
\ee
where the Levi-Civita connection $\tom^i$ is determined by the
condition $d\tom^i+\ve^i{}_{jk}\tom^j b^k=0$:
\bea
&&\tom^0=Nd\vphi\, ,\qquad \tom^1=-N^{-1}N_\vphi dr\, ,    \nn\\
&&\tom^2=-\frac{r}{\ell^2}dt+rN_\vphi d\vphi\, .           \lab{3.2c}
\eea
\esubeq
The pair $(b^i,\om^i)$ in \eq{3.2} defines the Euclidean black hole
in Riemann-Cartan spacetime.

Let us also display here the Euclidean \ads3\ metric, which is formally
obtained from \eq{3.1} by the replacements $8Gm=-1$, $J=0$, and
$\vphi=\phi$:
\be
ds^2=f^2dt^2+f^{-2}dr^2+r^2d\phi^2\, ,
     \qquad f^2\equiv 1+\frac{r^2}{\ell^2}\, .             \lab{3.3}
\ee
The same replacement in \eq{3.2} yields \ads3\ with torsion. Using
$\phi$ instead of $\varphi$, we wish to stress the difference in
topological properties between the black hole and \ads3, as discussed
in Appendix B.

\section{Energy and angular momentum}

For isolated macroscopic systems, energy and angular momentum are
dynamical quantities of fundamental importance for their thermodynamic
behavior. With the inclusion of gravity, these quantities can be
expressed as certain surface integrals over the asymptotic values of
dynamical variables. As a first step in our approach to the
thermodynamics of black holes with torsion, we use the standard
canonical formalism to calculate energy and angular momentum as the
asymptotic charges of the Euclidean black hole \eq{3.2}.

\subsection{Asymptotic conditions}

For any gauge theory, asymptotic conditions and their symmetries are
of essential importance for the physical content of the theory, as
they give rise to the conserved charges, which characterize the
dynamical behavior of the system. General asymptotic structure of the
Minkowskian 3D gravity with torsion and its relation to conformal
symmetry is well understood \cite{18}. Here, we wish to find global
charges of the Euclidean black hole, which is a much simpler problem.

The canonical procedure for calculating global charges is well
established.  We start by choosing the asymptotic conditions at
spatial infinity so that the fields $b^i$ and $\om^i$ are restricted
to the family of black hole configurations \eq{3.2}, parametrized by
$m$ and $J$. In other words, $b^i$ and $\om^i$ have the following
behavior as $r\to\infty$:
\bsubeq\lab{4.1}
\bea
&&b^i{_\m}\sim\left( \ba{ccc}
   \dis\frac{r}{\ell}-\frac{4Gm\ell}{r}    & 0  & 0   \\
   0 & \dis\frac{\ell}{r}+\frac{4Gm\ell^3}{r^3} & 0   \\
   -\dis\frac{4GJ}{r}                      & 0  & r
                     \ea
              \right) \, ,                                 \lab{4.1a}\\
&&\om^i{_\m}\sim\left( \ba{ccc}
   0 & 0 & \dis\frac{r}{\ell}-\frac{4Gm\ell}{r}       \\
   0 & \dis\frac{4GJ\ell}{r^3} &  0                   \\
   \dis-\frac{r}{\ell^2} & 0 & -\dis\frac{4GJ}{r}
                       \ea
                \right)+\frac{p}{2}\,b^i{_\m}\, .          \lab{4.1b}
\eea
\esubeq
Since these conditions represent a {\it restricted\/} version of the
general anti-de Sitter asymptotics, they are sufficient to define
only a restricted set of the conformal charges---energy and angular
momentum \cite{5,18}.

Having chosen the asymptotic conditions, we now wish to find the subset
of gauge transformations \eq{2.2} that respect these conditions and
define the asymptotic symmetry. They are determined by restricting the
original gauge parameters in accordance with \eq{4.1}, which yields
\be
\xi^\m=(\ell T_0,0,S_0)\, , \qquad   \th^i=(0,0,0)\, ,     \lab{4.2}
\ee
where $T_0$ and $S_0$ are arbitrary constants. In other words, the
asymptotic symmetry is described by two Killing vectors, $\pd/\pd t$
and $\pd/\pd\vphi$, as could have been concluded directly from the
form of the black hole solution \eq{3.2}. The corresponding
asymptotic symmetry group is $SO(2)\times SO(2)$, a subgroup of the
conformal group in two dimensions, in accordance with our choice of
the asymptotic conditions.

\subsection{Asymptotic charges}

As we shall see, the asymptotic conditions \eq{4.1} are acceptable at
the canonical level, since the related asymptotic symmetry has
well-defined canonical generators. The construction of the improved
generator and the corresponding asymptotic charges follows the
standard canonical procedure \cite{23,24,18}.

\prg{Hamiltonian and constraints.} Introducing the canonical momenta
$(\pi_i{^\m},\Pi_i{^\m})$, corresponding to the Lagrangian variables
$(b^i{_\m},\om^i{_\m})$, the action \eq{2.4} leads to the following
primary constraints:
\bea
&&\phi_i{^0}\equiv\pi_i{^0}\approx 0,\quad
  \phi_i{^\a}\equiv\pi_i{^\a}-\a_4\ve^{0\a\b}b_{i\b}\approx 0,\nn\\
&&\Phi_i{^0}\equiv\Pi_i{^0}\approx 0,\quad
  \Phi_i{^\a}\equiv\Pi_i{^\a}
  -\ve^{0\a\b}(2ab_{i\b}+\a_3\om_{i\b})\approx 0.          \nn
\eea
The canonical Hamiltonian is linear in unphysical variables:
\bea
&&\cH_c=b^i{_0}\cH_i+\om^i{_0}\cK_i+\pd_\a D^\a\, ,        \nn\\
&&\cH_i=-\ve^{0\a\b}\left(aR_{i\a\b}+\a_4T_{i\a\b}
   -\L \ve_{ijk}b^j{_\a}b^k{_\b}\right) \, ,               \nn\\
&&\cK_i=-\ve^{0\a\b}\left(aT_{i\a\b}+\a_3R_{i\a\b}
   +\a_4\ve_{imn}b^m{_\a}b^n{_\b}\right)\, ,               \nn\\
&& D^\a=\ve^{0\a\b}\left[\om^i{_0}(2ab_{i\b}+\a_3\om_{i\b})
                         +\a_4b^i{_0}b_{i\b}\right]\, .    \nn
\eea
In gauge theories, general dynamical evolution is governed by the
total Hamiltonian, which is obtained from $\cH_c$ by adding a linear
combination of the primary constraints. Thus,
$$
\cH_T=\cH_c+u^i{_\m}\phi_i{^\mu}+v^i{_\mu}\Phi_i{^\mu}\, ,
$$
where $u^i{_\m}$ and $v^i{_\mu}$ are arbitrary Hamiltonian
multipliers. The consistency conditions on the constraints lead to
the determination of $u^i{_\a}$ and $v^i{_\a}$, whereupon $\cH_T$
takes its final form:
\bsubeq\lab{4.3}
\bea
&&\cH_T=\hcH_T+\pd_\a\bD^\a\, ,                            \nn\\
&&\hcH_T=b^i{_0}\bcH_i+\om^i{_0}\bcK_i
         +u^i{_0}\pi_i{^0}+v^i{_0}\Pi_i{^0} \, ,           \lab{4.3a}
\eea
where
\bea
&&\bcH_i=\cH_i-\nabla_\b\phi_i{^\b}
   +\ve_{imn}b^m{_\b}\left(p\phi^{n\b}+q\Phi^{n\b}\right)\,,\nn\\
&&\bcK_i=\cK_i
   -\nabla_\b\Phi_i{^\b}-\ve_{imn}b^m{_\b}\phi^{n\b}\, .   \nn\\
&&\bD^\a=D^\a+b^i{_0}\phi_i{^\a}+\om^i{_0}\Phi_i{^\a}
        =b^i{_0}\pi_i{^\a}+\om^i{_0}\Pi_i{^\a} \, .        \lab{4.3b}
\eea
\esubeq
The constraints ($\pi_i{^0},\Pi_i{^0},\bcH_i,\bcK_i$) are first
class, while ($\phi_i{^\a},\Phi_i{^\a}$) are second class.

\prg{Canonical generator.} Applying the general Castellani's
algorithm \cite{24}, the canonical gauge generator is expressed in
terms of the first class constraints as follows:
\bea
G&=&-G_1-G_2\, ,                                           \lab{4.4}\\
G_1&\equiv&\dot\xi^\r\left(b^i{_\r}\pi_i{^0}
  +\om^i{_\r}\Pi_i{^0}\right)
  +\xi^\r\left[b^i{_\r}\bcH_i +\om^i{_\r}\bcK_i
  +(\pd_\r b^i{_0})\pi_i{^0}+(\pd_\r\om^i{_0})\Pi_i{^0}\right]\,,\nn\\
G_2&\equiv&\dot\th^i\Pi_i{^0}
  +\th^i\left[\bcK_i-\ve_{ijk}\left( b^j{_0}\pi^{k0}
  +\om^j{_0}\,\Pi^{k0}\right)\right]\, .                   \nn
\eea
Here, the time derivatives $\dot b^i{_0}$ and $\dot\om^i{_0}$ are
shorts for $u^i{_0}$ and $v^i{_0}$, respectively, and the
integration symbol $\int d^2x$ is omitted for simplicity. The
transformation law of the fields, $\d_0\phi\equiv\{\phi\,,G\}$, is
in complete agreement with the gauge transformations \eq{2.2} {\it on
shell\/}.

The behaviour of the momentum variables at large distances is
determined by the following ge\-ne\-ral principle: the expressions
that vanish on-shell should have an arbitrarily fast asymptotic
decrease, as no solution of the field equations is thereby lost.
Applying this principle to the primary constraints, we find the
asymptotic behavior of $\pi_i{^\m}$ and $\Pi_i{^\m}$.

The canonical generator acts on functions of the phase-space variables
via the Poisson bracket operation, which is defined in terms of
functional derivatives. In general, $G$ does not have well-defined
functional derivatives, but this can be corrected by adding suitable
{\it surface terms\/}. The improved canonical generator $\tG$ is found
to have the following form:
\bea
\tG=G+\G\, ,&&\qquad
      \G=-\int_0^{2\pi}d\vphi
          \left(\xi^0\cE^1+\xi^2\cM^1\right)\, ,           \lab{4.5}\\
\cE^\a\equiv
  2\ve^{0\a\b}&&\left[\left(a+\frac{\a_3p}{2}\right)\om^0{}_\b
  +\left(\a_4+\frac{ap}{2}\right)b^0{}_\b\frac{a}{\ell}b^2{}_\b
  -\frac{\a_3}{\ell}\om^2{}_\b\right]b^0{}_0\, ,           \nn\\
\cM^\a\equiv
  2\ve^{0\a\b}&&\left[\left(a+\frac{\a_3p}{2}\right)\om^2{}_\b
  +\left(\a_4+\frac{ap}{2}\right)b^2{}_\b
  +\frac{a}{\ell}b^0{}_\b
  +\frac{\a_3}{\ell}\om^0{}_\b\right]b^2{}_2\, ,           \nn
\eea
where $\xi^\m$ are the asymptotic parameters \eq{4.2}, i.e.
constants. The adopted asymptotic conditions guarantee
differentiability and finiteness of $\tG$; moreover, $\tG$ is also
{\it conserved\/}.

\prg{Canonical charges.} The value of the improved generator $\tG$
defines the {\it asymptotic charges\/}. Since $\tG\approx \G$, the
charges are completely determined by the boundary term $\G$.
Canonical expressions for the energy and angular momentum are defined
as the values of the surface term $-\G$, calculated for $\xi^0=1$ and
$\xi^2=1$, respectively. However, what we really need are the charges
corresponding to the analytically continued action $\bI=-I_E$, which
introduces an additional minus sign:
\be
E=-\int_0^{2\pi}d\vphi\,\cE^1\, ,\qquad
M=-\int_0^{2\pi}d\vphi\,\cM^1\, .                          \lab{4.6}
\ee
Consequently, energy and angular momentum of the black hole are given
by
\be
E= m+\frac{\a_3}{a}\left(\frac{pm}{2}-\frac{J}{\ell^2}\right),
\quad M= J+\frac{\a_3}{a}\left(\frac{pJ}{2}+m\right).      \lab{4.7}
\ee
Thus, the conserved charges are linear combinations of $m$ and $J$.
The expressions \eq{4.7} generalize the well-known results for the
conserved charges in \grl\ (where $E=m$ and $M=J$), and give us a new
physical interpretation of the parameters $m$ and $J$. Note also that
transition to Riemannian theory ($p=0$) still yields a non-trivial
modification of the \grl\ result \cite{19}.

\section{The black hole entropy}

Thermodynamic properties of black holes are closely related to the
quantum nature of gravity. In this section, we shall examine the role
of torsion in quantum 3D gravity by studying the partition function
of the Euclidean 3D gravity with torsion.

\prg{Partition function.} Let us consider the functional
integral (in units $\hbar=1$)
\bsubeq
\be
Z[\b,\Om]=\int DbD\om\exp\left(-\tI[b,\om,\b,\Om]\right)\,,\lab{5.1a}
\ee
where $\b$ and $\Om$ are the Euclidean time period and the angular
velocity of the black hole, respectively, the fields $(b^i,\om^i)$
satisfy certain boundary conditions, and $\tI$ is the Euclidean
action \eq{2.4}, corrected by suitable boundary terms. The {\it
Lagrangian\/} boundary conditions define a set of the allowed field
configurations $\cC_L$, such that: \\
(i) $\cC_L$ contains black holes with $(m,J)$ belonging to a small
    region around some fixed $(m,J)_0$, \\
(ii) $\b$ and $\Om$ remain constant on the boundary, and \\
(iii) one can construct the boundary terms which make the improved
action $\tI$ differentiable.

If we recall the definition of $\b$ and $\Om$ from section III, we
can see that they are determined in terms of $r_+$ and $\r_-$
(functions of $m$ and $J$), so that the conditions (i) and (ii) may
become incompatible. However, this is not a sincere problem. Indeed,
it can be resolved by treating $\b$ and $\Om$ as {\it independent\/}
parameters, determined in terms of $r_+$ and $\r_-$ only ``on shell".
Geometrically, an ``off-shell" extension of $\b$ and $\Om$ leads to
conical singularities, but at the end, after imposing the field
equations, they disappear \cite{8,9}. Interpreted in this way, the
above boundary conditions correspond to the {\it grand canonical
ensemble\/}.

Although the partition function \eq{5.1a} cannot be calculated
exactly, the semiclassical approximation around the black hole
solution \eq{3.2} gives very interesting insights into the role of
torsion in quantum dynamics. Indeed, starting with the semiclassical
expansion $\tI=\tI_{\rm bh}+\cO(\hbar)$, where $\tI_{\rm bh}$ is the
value of the classical action $\tI$ at the black hole configuration,
one finds, to the lowest order in $\hbar$, that the partition
function is given by $\ln Z[\b,\Om]\approx -\tI_{\rm bh}$. This
result should be compared with the general form of the grand
canonical partition function:
\be
Z[\bar\b,\mu]=e^{-\bar\b F}\, ,\qquad F=E-\mu C-TS\, ,     \lab{5.1b}
\ee
where $F$ is the free energy, $\bar\b=1/T$ is the inverse
temperature, $E$ and $S$ are the energy and entropy of the system,
and $\mu$ is the chemical potential corresponding to the conserved
charge $C$. Since the classical values of $E$ and the (second)
conserved charge $C=M$ are already known from the canonical analysis,
it follows that the classical action alone is sufficient to give us a
non-trivial information on the black hole entropy:
\be
\tI_{\rm bh}=\bar\b(E-\mu C)-S\, .                         \lab{5.1c}
\ee
\esubeq

\prg{The canonical action.} Instead of working directly with the
covariant action \eq{2.4}, we shall rather use its canonical form, as
in \cite{3}:
\bsubeq
\be
\Ic=\int dt\int d^2x\left(\pi_i{^\m}\dot{b}^i{_\m}
                    +\Pi_i{^\m}\dot{\om}^i{_\m}-\hcH_T\right)\, ,
\ee
where $\hcH_T$ is the total Hamiltonian \eq{4.3}. The Lagrangian
boundary conditions (i), (ii) and (iii), can be easily extended to
the {\it Hamiltonian\/} boundary conditions, defined by the related
phase-space configurations $\cC_H$. The action $\Ic$ does not have
well-defined functional derivatives, since its variation on $\cC_H$
produces not only the field equations, but also some boundary terms.
The improved action has the general form
\be
\tIc=\Ic+B\, ,                                             \lab{5.2b}
\ee
\esubeq
where the boundary term $B$ is chosen to make $\tIc$ differentiable.
The partition function is now given as the functional integral over
$\cC_H$.

For a thermodynamic system in equilibrium, the ensemble must be time
independent. If one calculates the value of $\tIc$ on the set of
static triads and connections, one finds that $\Ic$ vanishes on shell
($\hcH_T\approx 0$), so that the only term that remains is the
boundary term $B$. This term contains the complete information about
the black hole thermodynamics.

\subsection{Boundary terms}

The boundary term $B$ is constructed by demanding $\d\Ic+\d B\approx
0$, where $\approx$ denotes an equality when the Hamiltonian
constraints hold (on-shell or weak quality). In other words, $B$
should cancel the unwanted boundary terms in $\d\Ic$, arising from
integrations by parts.

Using the relation $\d\hcH_T\approx b^i{_0}\d\bcH_i+\om^i{_0}\d\bcK_i$,
we find that the general variation of $\Ic$ at fixed $r$ has the form
\bea
\d\Ic\big|^r&=&-\int dt\int d^2 x\d\cH_T\Big|^r            \nn\\
  &\approx& 2\int dtd\vphi\left[b^i{_0}(a\d \om_{i2}
   +\a_4\d b_{i2}) +\om^i_0(a\d b_{i2}
   +\a_3\d\om_{i2})\right]^r\, .                           \nn
\eea
We can now restrict the phase space by the requirement
$$
\om^i=\tom^i+\frac{p}{2}b^i+\hcO\, ,
$$
where $\hcO$ is arbitrarily small at the boundary, as no solution of
the field equations is thereby lost. Consequently,
\bea
\d\Ic\big|^r&\approx& 2\int dtd\vphi
   \left[-\frac{\a_3}{\ell^2}b^i{_0}\d b_{i2}
   +a(b^i{_0}\d\tom_{i2}+\tom^i{_0}\d b_{i2})\right.       \nn\\
&&\left.+\a_3\tom^i{_0}\d\tom_{i2}
   +\a_3\frac{p}{2}\left(b^i{_0}\d\tom_{i2}
   +\tom^i{_0}\d b_{i2}\right)\right]^r\, .                \lab{5.3}
\eea
The boundary term $B$ contains the contributions from infinity and
from the horizon, which are determined by the requirement
$$
\d I_c\big|^{r\to\infty}-\d I_c\big|^{r_+}
                        +\d(B^\infty+B^{r_+})\approx 0\, .
$$

\prg{Spatial infinity.} The boundary term stemming from infinity has
been already calculated in the construction of the improved
Hamiltonian. It has the form
\bea
&&\d I_c\big|^{r\to\infty}
    =-\int_0^\b dt\left[\d\int d^2x\,\hcH_T\right]^{r\to\infty}
   \approx-\d B^\infty\, ,                                 \nn\\
&&B^\infty=\b E\, .                                        \lab{5.4}
\eea
Here, $E$ is the canonical energy \eq{4.7}, and the time period $\b$
is kept fixed, since we are in the grand canonical ensemble.

\prg{The horizon.} If we consider Minkowskian black hole as a
macroscopic object, it seems quite natural to treat only its ``outer"
part $r>r_+$ as physical. The consistency of this idea leads to
certain boundary conditions at $r=r_+$, which give rise to additional
boundary terms \cite{8,22} (see also \cite{25}). In the Euclidean
formalism, in spite of the fact that the ``inner" region $r<r_+$ is
absent from the black hole manifold, there are convincing arguments
that one still needs boundary terms at the horizon \cite{8,21,22}.
These arguments are based on the observation that the Killing vector
field $\pd_t$ is not well defined at $r=r_+$. Motivated by these
considerations, we assume that the line $r=r_+$ is removed from the
black hole manifold, which modifies the topology: the horizon becomes
an additional (inner) boundary of the manifold.

After accepting such an assumption, we have to introduce appropriate
boundary conditions at the horizon, in order to further improve the
differentiability of the action. Let us observe that the values of
the triad field and Riemannian connection satisfy the following
relations on the horizon:
\be
b^a{_0}-\Om b^a{_2}=0\, ,\qquad
\tom^a{_0}-\Om\tom^a{_2}=-\frac{2\pi}{\b}\d^a_2\, ,        \lab{5.5}
\ee
for $a=0,2$. Having in mind that the boundary conditions should
correspond to the grand canonical ensemble, we promote these
``on-shell'' relations into the ``off-shell" boundary conditions at
the horizon, with $\Om$ and $\b$ as independent parameters (compare
with \cite{8,18}).

Now, we are ready to calculate the corresponding boundary term. Using
the general relation \eq{5.3} and the boundary conditions \eq{5.5},
we find
\bea
\d\Ic\big|^{r_+}&\approx&\d B^{r_+}\, ,                    \nn\\
B^{r_+}&=&2\int_0^\b dt\int_0^{2\pi}d\phi
  \Bigg\{\Om\left[-\frac{\a_3}{2\ell^2} b^i{_2}b_{i2}\right.\nn\\
&&+\left.\left(a+\frac{\a_3p}{2}\right)b^i{_2}\tom_{i2}
  +\frac{\a_3}{2}\tom^i{_2}\tom_{i2}\right]
  -\frac{2\pi}{\b}\left[\left(a+\frac{\a_3p}{2}\right)b_{22}
  +\a_3\tom_{22}\right]\Bigg\} \, .                        \nn
\eea
In order to calculate $B^{r_+}$ at the black hole configuration, we
use the relations
$$
b^i{_2}b_{i2}\big|^{r_+}=r_+^2\, ,\quad
b^i{_2}\tom_{i2}\big|^{r_+}=-4GJ\, ,\quad
\tom^i{_2}\tom_{i2}\big|^{r_+}=\frac{\r_-^2}{\ell^2}\, ,
$$
which yield
\be
B^{r_+}=-\b\Om M-\left[\frac{2\pi r_+}{4G}
   +4\pi^2\a_3\left(pr_+-2\frac{\r_-}{\ell}\right)\right]\, .\lab{5.6}
\ee

\subsection{Entropy}

The value of the improved action at the black hole configuration is
equal to the sum of boundary terms \eq{5.4} and \eq{5.6}:
\bea
\tI_{\rm bh}&=&B^{\infty}+B^{r_+}                          \nn\\
  &=&\b(E-\Om M)-\left[\frac{2\pi r_+}{4G}
    +4\pi^2\a_3\left(pr_+-2\frac{\r_-}{\ell}\right)\right].\lab{5.7}
\eea
The thermodynamic interpretation of this result, based on Eq.
\eq{5.1c}, tells us that $\b$ is the inverse temperature, $\Om$ is
the chemical potential corresponding to $M$, and
\be
S=\frac{2\pi r_+}{4G}
  +4\pi^2\a_3\left(pr_+-2\frac{\r_-}{\ell}\right)          \lab{5.8}
\ee
is the entropy of the black hole with torsion.

Microscopic interpretation of the black hole entropy is naturally
related to the number of dynamical degrees of freedom located at the
boundary of spacetime \cite{4,8,9}. For $\a_3=0$, our result \eq{5.3}
reduces to the first term---the Bekenstein-Hawking value of $S$. The
additional term, proportional to $\a_3$, stems from the Chern-Simons
contribution to the action \eq{2.4}. The first piece of this term,
proportional to $pr_+$, can be interpreted as the contribution of the
torsion degrees of freedom at the outer horizon, while the second
piece is due to degrees of freedom at the ``inner horizon" $\r_-$.
Since the Euclidean black hole manifold does not contain the inner
horizon ($r_-=-i\r_-$ is imaginary), the appearance of $\r_-$ in
(5.8) could be understood as a consequence of the analytic structure
of the theory, which is expected to contain relevant information on
the Minkowskian sector. For vanishing torsion ($p=0$), the entropy
reduces to the result obtained by Solodukhin, in his study of
Riemannian \grl\ with a Chern-Simons term \cite{19}.

For a given black hole with fixed $r_+$ and $\r_-$, the condition $S\ge
0$ imposes certain bounds on the parameters $\a_3$, $p$ and $\ell$. It
is an interesting question what happens with $S$ at the absolute zero
of temperature, $T=(r_+^2+\r_-^2)/2\pi\ell^2r_+\to 0$. Formally, the
black hole at the absolute zero is in the ground state, defined by
$r_+,\r_-\to0$ ($m,J\to 0$), and the entropy \eq{5.8} of the ground
state vanishes, in agreement with the third law of thermodynamics.
However, this line of arguments is not acceptable, since in the ground
state region, the semiclassical result \eq{5.8} is outside of its
domain of validity. Indeed, for $S=0$, the generalized Smarr formula
$2\b(E-\Om M)=S$ (obtained by direct calculation) implies that the
whole $\tI_{\rm bh}$ vanishes; but if $\tI_{\rm bh}=0$, the 0-loop
approximation of the semiclassical expansion is not reliable. Thus, the
mathematical limitations of the result \eq{5.8} do not allow us to have
a true estimate of the black hole entropy in the extreme case of the
black hole ground state.

Using the rules of Euclidean continuation described in Appendix A,
the entropy can be easily expressed in terms of the corresponding
Minkowskian parameters \cite{18}:
\be
S=\frac{2\pi r_+}{4G}
  +4\pi^2\a_3\left(pr_+-2\frac{r_-}{\ell}\right)\, .       \lab{5.9}
\ee

\subsection{The first law of thermodynamics}

If the black hole solution is an extremum of the canonical action
$\Ic$ on the set of allowed phase-space configurations, $\d
\Ic\vert_{\rm bh}\approx-\d(B^\infty+B^{r_+})\vert_{\rm bh}=0$, we
obtain the relation
\be
\d S=\b\d E-\b\Om\d M\, ,                                  \lab{5.10}
\ee
which represents the first law of black hole thermodynamics (compare
with arguments given in Ref. \cite{25}). The result \eq{5.10} is also
confirmed by a direct calculation, using our expressions for $S,E$
and $M$. Thus, the existence of torsion in 3D is in complete
agreement with the first law of black hole thermodynamics.

Using $\b=1/T$, the first law \eq{5.10} can be written in the form
\be
T\d S= \d E-\Om\d M\, .                                    \lab{5.11}
\ee

\section{Concluding remarks}

In this paper, we investigated the role of torsion in the black hole
thermodynamics by studying the grand canonical partition function of
the Euclidean black hole with torsion, in the lowest-order
semiclassical approximation.

(1)  The black hole entropy is obtained from the boundary term at the
horizon.

(2) It differs from the Bekenstein-Hawking result by an additional
term, which describes the torsion degrees of freedom at the outer
horizon, and degrees of freedom at the ``inner horizon". For $p=0$,
we have a Riemannian theory with Chern-Simons term in the action, and
our $S$ coincides with Solodukhin's result \cite{19}.

(3) The existence of torsion is in complete agreement with the
first law of black hole thermodynamics.

\acknowledgements

We would like to thank Friedrich Hehl for a careful reading of the
manuscript and a number of enlightening comments and suggestions.
This work was supported in part by the Serbian Science Foundation,
Serbia.

\appendix
\section{Euclidean continuation}

Euclidean continuation of Minkowskian black holes can be formally
expressed as a mapping $f_E$ from Minkowskian to Euclidean variables,
such that
\be
f_E:\quad t\mapsto -it\, ,\qquad J\mapsto -iJ \, .
\ee
As a consequence, the induced mapping of the triad field reads
$$
b^0\mapsto -ib^0\, ,\qquad b^1\mapsto b^1\, ,\qquad
b^2\mapsto b^2\, .
$$
The analytic continuation maps $\eta^M=(1,-1,-1)$ into
$\bar\eta=(-1,-1,-1)$. Note, however, that we define our Euclidean
theory to have the positive-definite metric:
$$
f_E:\quad \eta^M\mapsto\eta^E=(1,1,1)\, .
$$
Demanding that the torsion $T^i$ maps in the same way as $b^i$, we
find
$$
\om^0\mapsto-\om^0\, ,\qquad\om^1\mapsto -i\om^1\, ,\qquad
\om^2\mapsto-i\om^2\, ,
$$
which then defines the mapping of the curvature $R^i$. It is now easy
to derive the mappings of different terms in the Mielke-Baekler
action:
$$
\ba{l}
  b^iR_i\mapsto ib^iR_i\, , \\[3pt]
  \cL_\cs(\om)\mapsto\cL_\cs(\om)\, ,
\ea \qquad
\ba{l}
  b^0b^1b^2\mapsto -ib^0b^1b^2 \, , \\[3pt]
  b^iT_i\mapsto-b^iT_i\, ,
\ea
$$
where, on the right-hand sides, we are using not $\bar\eta$, but
$\eta^E$. Now, the mapping of the complete action integral,
\be
f_E:\quad iI\mapsto -I\, ,
\ee
can be effectively expressed by the following mapping of parameters:
\be
\ba{ll}
  f'_E:\quad & a\mapsto a\, , \\
             & \a_3\mapsto i\a_3\, ,
\ea \qquad
\ba{l}
  \L\mapsto-\L\, , \\
  \a_4\mapsto -i\a_4\, ,
\ea
\ee
In particular, using $-1/\ell^2=\Leff\mapsto-\Leff=-1/\ell^2$, we see
that $\Leff$ changes the sign, while $\ell$ remains unchanged.

\section{Hyperbolic 3D space}

Here, we review some facts about the Euclidean \ads3, known also as
the hyperbolic 3D space $H^3$ (see, e.g. \cite{9,26}). Consider a
hypersurface
$$
(y^0)^2 +(y^1)^2+(y^2)^2-z^2=-\ell^2\, ,\qquad \ell^2>0 \, ,
$$
embedded in a four-dimensional Minkowski space $M_4$ with metric
$\eta_{MN}=(1,1,1,-1)$. The hypersurface consists of two disjoint
hyperboloids, with $z\ge\ell$ and $z\le-\ell$, and $H^3$ can be
visualized as one of these hyperboloids. Clearly, $H^3$ has the
isometry group $SO(1,3)$, and can be thought of as the coset space
$SO(1,3)/SO(3)$. The Riemannian scalar curvature of $H^3$ is
negative, $\tR=-6/\ell^2$, and the signature is $(+,+,+)$.

Using the parametrization
$$
\ba{l}
  y^0=\ell\cosh\r\sinh\psi\, ,\\
  z=\ell\cosh\r\cosh\psi\, ,
\ea \qquad
\ba{l}
  y^1=\ell\sinh\r\cos\phi\, ,\\
  y^2=\ell\sinh\r\sin\phi\, .
\ea
$$
the metric on $H^3$ ($z\ge\ell$) takes the form
\be
ds^2=\ell^2\left(d\r^2+\cosh^2\r\,d\psi^2+\sinh^2\r\,d\phi^2\right)\,.
\ee
The change of coordinates $\psi=t/\ell$, $r=\ell\sinh\r$, shows that
$H^3$ is isometric to the Euclidean version of \ads3, equation
\eq{3.3}.

Introducing $\cos\chi=1/\cosh\r$, we find another useful form of the
metric:
\be
ds^2=\frac{\ell^2}{\cos^2\chi}
      \left(d\chi^2+d\psi^2+\sin^2\chi d\phi^2 \right)\, . \lab{B2}
\ee

The coordinate transformation
\bea
&&\phi=\frac{r_+}{\ell^2}t+\frac{\r_-}{\ell}\vphi\, ,\qquad
  \psi=\frac{r_+}{\ell}\vphi-\frac{\r_-}{\ell^2}t\, ,        \nn\\
&&\cos\chi=\sqrt{\frac{r_+^2+\r_-^2}{r^2+\r_-^2}}\, ,
\eea
yields the black hole metric in Schwarzschild coordinates
$(t,r,\vphi)$. The BTZ metric \eq{3.1} is obtained from the \ads3\
metric \eq{B2} by making the following (isometric) identifications:
\bitem
\item[(i)] $(\chi,\psi,\phi)\to(\chi,\psi,\phi+2\pi)$, which eliminates
the conical singularity at $\chi=0$ (horizon) in \eq{B2}, and is
equivalent to $(r,\vphi,t)\to(r,\vphi+\Phi,t+\b)$, with
$$
\Phi=\frac{2\pi\ell\r_-}{r_+^2+\r_-^2}\, ,\qquad
\b=\frac{2\pi\ell^2 r_+}{r_+^2+\r_-^2}\, .
$$\vsm
\item[(ii)] $(\chi,\psi,\phi)\to(\chi,\psi+2\pi r_+/\ell,
\phi+2\pi\r_-/\ell)$, which is equivalent to $\vphi\to\vphi+2\pi$.
\eitem
Thus, the Euclidean black hole may be described as the quotient of
the hyperbolic space $H^3$ by the isometry (i)+(ii). The topology of
the black hole manifold is a solid torus, $R^2\times S^1$.

The identification (i) shows that $\vphi$ is not the usual
Schwarzschild azimuthal angle $\vphi'$. The relation between them is
$\vphi'=\vphi+\Om t$, where $\Om=-\r_-/r_+\ell=N_\vphi(r_+)$. Indeed,
$$
(\vphi',t)\to(\vphi'+\Phi+\Om\b,t+\b)=(\vphi',t+\b)\, .
$$
Note that $d\vphi+N_\vphi dt=d\vphi'+(N_\vphi-\Om)dt$, so that
$N'_\vphi=N_\vphi-\Om=0$ at the horizon.

The Poincar\'e upper half-space model for $H^3$ is given by the
metric
\bea
ds^2=\frac{1}{z^2}\left(dx^2+dy^2+dz^2\right)\, ,\qquad z>0\, .
\eea
It follows from \eq{B2} by a simple coordinate transformation:
\bea
&&x=\exp\psi\,\sin\chi\cos\phi\, ,                         \nn\\
&&y=\exp\psi\,\sin\chi\sin\phi\, ,                         \nn\\
&&z=\exp\psi\,\cos\chi\, .                                 \nn
\eea

In the standard spherical coordinates with $R=\exp\psi$, we have:
\be
ds^2=\frac{\ell^2}{\cos^2\chi}
     \left(\frac{dR^2}{R^2}+d\psi^2+\sin^2\chi d\phi^2\right)\,.
\ee
The identification $\vphi\to\vphi+2\pi$ is now described by
$(\chi,R,\phi)\to(\chi,Re^{2\pi r_+/\ell},\phi+2\pi\r_-/\ell)$.


\begin{thebibliography}{99}
\bibitem[*]{*} The field equations are algebraic equations for
the field strengths $(T^i,R^i)$. In the degenerate case
$\a_3\a_4-a^2=0$, the number of independent equations reduces from six
to three, and the general solution for $(T^i,R^i)$ contains three
arbitrary functions. For a particular choice of these functions, one
still finds the black hole as an exact solution. Here, we restrict our
attention to the non-degenerate case in order to avoid dynamically
indeterminate situations.
\end{thebibliography}
\end{document}